\documentclass[pre,reprint,twocolumn,amsmath,10pt,amssymb,aps,superscriptaddress,showkeys,showpacs]{revtex4-1}\input epsf
\usepackage{bm}
\usepackage{graphicx}
\usepackage{makeidx}
\usepackage{chemarr}
 \usepackage{color}
 \definecolor{darkgreen}{rgb}{0,0.6,0}
 \definecolor{orange}{rgb}{0.99,0.257,0}
\newcommand{\be}{\begin{equation}} 
\newcommand{\ee}{\end{equation}}
\newcommand{\ba}{\begin{eqnarray}} 
\newcommand{\ea}{\end{eqnarray}}

\def\ie{{\it i.e}~}

\def\epsilon{\varepsilon}

\def\ie{i.e.}

\def\beqr{\begin{eqnarray}}
\def\eqnr{\end{eqnarray}}
\def\beq{\begin{equation}}
\def\bc{\begin{center}}
\def\ec{\end{center}}
\def\eqn{\end{equation}}

\usepackage{float}

\def\be{\begin{equation}}
\def\ee{\end{equation}}
\def\bea{\begin{eqnarray}}
\def\eea{\end{eqnarray}}

\usepackage{amsmath}

\usepackage{graphicx}
\usepackage{dcolumn}
\usepackage{bm}

\begin{document}
\title{Capture of a diffusive prey by multiple predators in confined space}
\author{Indrani Nayak}
\email{inayak21@phy.iitb.ac.in}
\author{Amitabha Nandi}
\email{amitabha@phy.iitb.ac.in}
\author{Dibyendu Das}
\email{dibyendu@phy.iitb.ac.in}
\affiliation{Department of Physics, Indian Institute of  Technology Bombay, Powai, Mumbai, 400076, India.}

\begin{abstract}
The first passage search of a diffusing target (prey) by multiple searchers (predators) in \emph{confinement} is an important problem in the stochastic process literature. While the analogous problem in open space has been studied in some details, a systematic study in confined space is still lacking. In this paper, we study the first passage times for this problem in 1,2 and $3-$dimensions. Due to confinement, the survival probability of the target takes a form $\sim e^{-t/\tau}$ at large times $t$. The \emph{characteristic} capture timescale $\tau$ associated with the rare capture events are rather challenging to measure. We use a computational algorithm that allows us to estimate $\tau$ with high accuracy. We study in details the behavior of $\tau$ as a function of the system parameters, namely, the number of searchers $N$, the relative diffusivity $r$ of the target with respect to the searcher, and the system size. We find that $\tau$ deviates from the $\sim 1/N$ scaling seen in the case of a static target, and this deviation varies continuously with $r$ and the spatial dimensions.

\begin{description}
\item[PACS number(s)]05.40.-a,02.50.-r,02.50.Ey,02.70.-c
\end{description}

\end{abstract}

\maketitle

\section{Introduction}

Search and capture processes are ubiquitous in nature and have been an important topic in stochastic processes \cite{redner,bray,ralf2014first}. Such processes find application in a wide range of fields \citep{chandrasekhar1989selected,viswanathan1999optimizing,*edwards2007revisiting, benichou2010optimal,rotbart2015michaelis,parmar2016theoretical,weng2017hunting,singh2019universal,nayak2020comparison, biswas2020first,belousov2020first,gomez2020target}, and the list is still growing.
Theoretical techniques to study a search and capture process involve the statistics of the first encounter with the target -- commonly known as the \emph{first passage times} (FPT). A complete characterization of an encounter problem is therefore possible by either studying the FPT distribution $F(t)$, or the probability of survival of the target $S(t)$, and the different moments and timescales associated with $F(t)$ (and $S(t)$)\cite{van1993short,condamin2005first,mejia2011first,godec2016universal,godec2016first,vot2020first}. \\

%
While a simple encounter process involves a single searcher-target pair, more realistic first passage problems may involve multiple targets and searchers. Trapping reactions are classic examples of such processes and studied extensively before \cite{bramson1988asymptotic,bray2002exact,oshanin2002trapping,mehra2002trapping,blythe2003survival,moreau2003pascal,*moreau2004lattice}. A variant of this problem is that of multiple walkers searching for a single target. In this case, the FPT denotes the time of the first encounter of any one of the entities with the target. The limiting case when the target is \emph{static} has been studied before in a variety of context \citep{van2003uphill,benichou2010geometry,mejia2011first,godec2016universal,godec2016first,ro2017parallel,hartich2018duality,basnayake2019asymptotic,lawley2020universal,*lawley2020distribution,grebenkov2020single} ---
for non-interacting searchers, the survival probability of the target takes the form $S(t)=[s_1(t)]^N$, where $s_1(t)$ is the survival probability in the presence of a single searcher \cite{mejia2011first}. While in \emph{free} space, mostly $s_1(t) \sim t^{-\gamma}$ as $t\to\infty$, in a confined space, $s_1(t)\sim\exp(-t/\tau)$ in the asymptotic limit \cite{mattos2014trajectory,godec2016first}. Here $\tau$ is the \emph{characteristic} time that represents the timescale associated with the rare events of capture. \\

What happens when the target itself is not static? This question leads to an interesting variant of the prey-predator type models \cite{krapivsky1996kinetics,redner1999capture,winkler2005drowsy,oshanin2009survival,gabel2012can,redner2014gradual,schwarzl2016single,toledo2019predator}. Commonly known as the \emph{lamb-lion} problem, here, a diffusive target (lamb) is chased by $N$ diffusive searchers (lions). The problem has been studied in great detail in free space \cite{krapivsky1996kinetics,redner1999capture,blythe2002perturbation,gabel2012can,grassberger2002go}. In $1-d$, $S(t)\sim t^{-\theta_N}$ asymptotically where in contrast to a static target, here $\theta_N$ not only depends on $N$ but also on the ratio of the diffusivities of the lamb and lion \cite{krapivsky1996kinetics,redner1999capture}. While the exact value of $\theta_N$ is known for $N=1,2$, for $N>2$, approximate calculations based on statistics of extremes show that  $\theta_N\sim\ln N$ \cite{krapivsky1996kinetics,redner1999capture}.  As $N\to\infty$, the dependency on $N$ is washed away, and $S(t)\sim\exp(-\ln^2t)$ \cite{krapivsky1996kinetics}.  The problem becomes difficult to solve analytically in higher dimensions.

While open space is relevant in many physical problems, on the other hand, there are many diffusive processes in nature that occur in confined spaces. Transport inside living cells are limited by the cellular dimensions --- relevant examples in this context are proteins binding to a target site on DNA \cite{berg1981diffusion} or microtubules trying to capture a kinetochore \cite{nayak2020comparison}. Likewise, neutrophils chase and engulf diffusing bacteria or foreign particles within a finite region of the bloodstream \cite{kikushima2013non}. Similarly in ecology, one may consider the movement of prey and predators confined within an island. Although most of these examples exhibit both active and passive transport, such examples serve as a natural motivation to study the classic lion-lamb problem under confinement for pure diffusion. Moreover, realistic examples are often in dimensions $d>1$, hence the dependence on dimensionality should be systematically studied. Note that for this problem, only a couple of analytical results are known, that too in $1-d$ and for a single searcher-target pair only. Assuming equal diffusivities of the lion-lamb pair, the full survival probability and hence the characteristic time are known exactly \cite{tejedor2011encounter}. For the case unequal diffusivities, a very recent study has shown that the characteristic time can be estimated approximately in the limit when the lamb diffuses much slower than the lion \cite{vot2020first}. For multiple lions, the problem is hard to solve analytically. In this paper, we take a computational approach to tackle the problem. Note that even computationally finding $S(t)$, particularly its asymptotic behavior characterized by $\tau$, is a challenging task. Often the asymptotic exponential tail is not visible unless one goes to extremely low values of $S(t)$. Here we use a numerical algorithm \cite{grassberger2002go}, which allows estimates of $S(t)$ to the  striking degree of accuracy ($\sim 10^{-100}$), thus allowing us to obtain $\tau$ unambiguously. While the mean first passage time is commonly used to characterize a capture process, previous studies have shown that mean times depend on the initial conditions, particularly for capture processes under confinement \cite{mejia2011first,mattos2012first}. On the other hand $\tau$ is independent of the initial positions of the walkers and hence is more robust \cite{nayak2020comparison}. Hence in this work, we solely focus on the behavior of $\tau$ as a function of the system parameters, namely the number of lions $N$, the relative diffusivity $r$ of the lamb with respect to a lion, and the dimensionality $d$. 


The structure of the paper is as follows. In Sec.~\ref{Model}, we introduce the model. In Sec.~\ref{Methods}, we discuss the numerical method to obtain $\tau$. In Sec.~\ref{Results}, we present the results for the variation of $\tau$ with the system parameters and a perspective based on  extreme value statistics related to the $N$ dependence of $\tau$. We discuss our results in Sec.\ref{conclusions}.

\section{Description of the model}
\label{Model}
\begin{figure}[!htbp]
\centering
\includegraphics[width=3.4in]{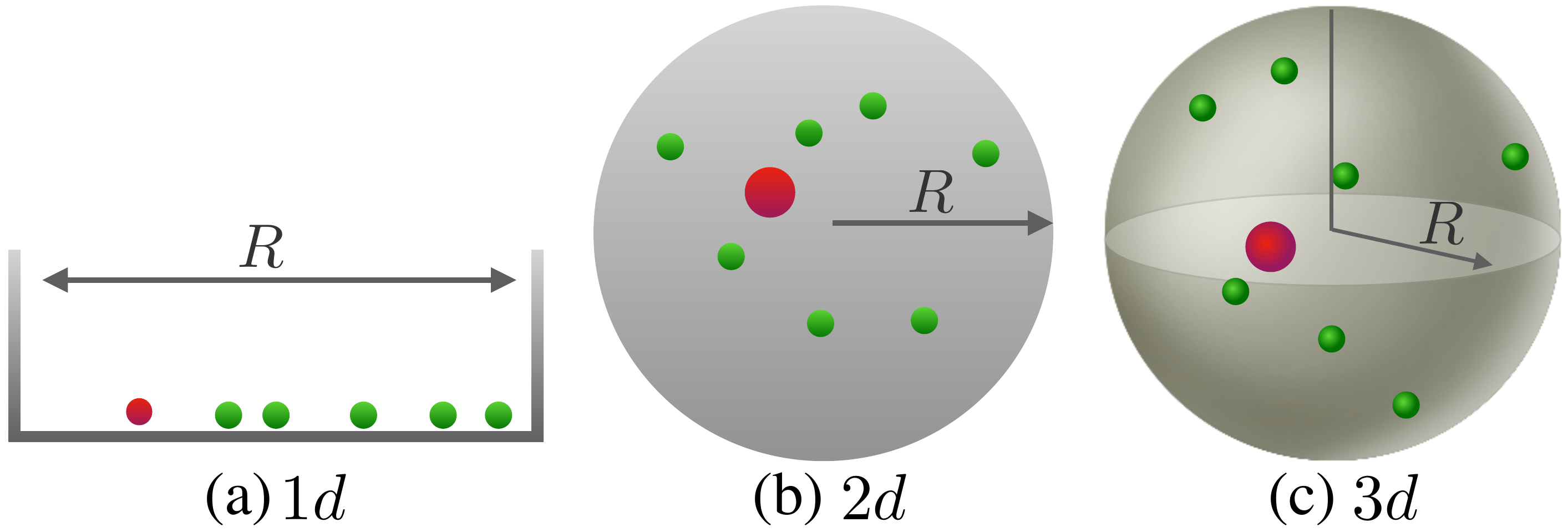} 
\caption{(Color online) Schematic of  $N$ non-interacting searchers or lions (green) are foraging for a diffusive target or the lamb (red) in  $d$-dimensional confined geometry. The lamb has diffusivity $D_l$, and all the lions have equal diffusivity $D_L$. The confined region is (a) a line segment of length $R$ in $1d$, (b) a circle  in $2d$, and (c) a sphere in $3d$ of radius $R$. The lamb is considered to be a circle and a sphere of radius $a$ in $d=2,3$ respectively. In $1d$, the system is bounded by two reflecting boundaries at $x=0$ and $R$.  In $2d$ and  $3d$, the system is bounded by a radially symmetric reflecting boundary at radius $R$. 
}
\label{Fig_models}
\end{figure}

Our system consists of a \emph{moving} target (lamb) with $N$ searchers (lions) in $d$-dimensional closed volume (see Fig.~\ref{Fig_models}).  In $1-d$, the bounding volume is a box with two reflecting walls at $x=0$ and $R$ [Fig.~\ref{Fig_models}(a)]. For $d=2$ and $d=3$, we chose the confining volumes to be a circle and a sphere, each of radius $R$ respectively [Fig.~\ref{Fig_models}(b),(c)]. The lions are non-interacting point particles with equal diffusivities $D_L$. The lamb has a diffusivity $D_l$ and is chosen to be a point particle in $1d$ [Fig.~\ref{Fig_models}(a)], and a circle or a sphere of radius $a$ in $d=2$ and $d=3$ (Fig.~\ref{Fig_models}(b),(c)). At $t=0$, the lamb is placed to the left of the lions in $1d$, and in $2-d$ and $3-d$, it is always placed at the centers of the circle and the sphere respectively. The lions start from the same initial point in all our simulations. Because of the choice of the boundary conditions and initial conditions, radial symmetry is ensured and the survival probability will only depend on the initial radial positions of the lions and time $t$. The condition of capture of the lamb by any of the lions in $d-$dimension is: $\mid{\mathbf{r}_i(t) -\mathbf{r}_l(t)}\mid \le a$, where $\mathbf{r}_l(t),\mathbf{r}_i(t)$ represent the positions of the lamb and the $i^{th}$ lion at any instant $t$. In $1-d$ since $a=0$, the capture condition is $(r_i(t) -r_l(t))\le 0$. Here we focus on the behavior of $\tau$ as a function of $N$, the relative diffusivities $r=D_l/D_L$ of lamb-lion, and the system size $R$ in $d=1,2,3$.

\par  The backward Fokker-Planck equation \cite{gardiner} for the survival probability of the lamb in the presence of $N$ lions is as follows:
\beq
\frac{\partial S(t,\mathbf{r}_l,\{\mathbf{r}_i\})}{\partial t}=D_l \nabla ^2_{l}S(t,\mathbf{r}_l,\{\mathbf{r}_i\})+D_L \sum_{i=1}^N \nabla ^2_{i}S(t,\mathbf{r}_l,\{\mathbf{r}_i\}).
\eqn
Here $S(t,\mathbf{r}_l,\{\mathbf{r}_i\})$ is the survival probability of the lamb up to time $t$ with the initial positions $\mathbf{r}_l$ and $\mathbf{r}_i$ of the lamb and the $i^{th}$ lion respectively. Here $\nabla^2$ is the $d-$dimensional Laplacian. Dividing the above equation by $D_L$, we get 
\beq
\frac{\partial S(t,\mathbf{r}_l,\{\mathbf{r}_i\})}{\partial (D_L t)}=r \nabla ^2_{l}S(t,\mathbf{r}_l,\{\mathbf{r}_i\})+ \sum_{i=1}^N \nabla ^2_{i}S(t,\mathbf{r}_l,\{\mathbf{r}_i\}).
\label{BFP}
\eqn
As commented earlier, although the survival probability depends on the initial positions $\mathbf{r}_l$ and $\{\mathbf{r}_i\}$'s, the characteristic times $\tau$ are independent of them and depend on the diffusivities $D_l$ and $D_L$. However in Eq.~(\ref{BFP}), the RHS is only dependent on the parameter $r$. This indicates that the scaled characteristic time $\tau D_L$ would not depend on $D_L$ and $D_l$ separately but only on their ratio $r$. This significantly simplifies the parameter landscape if one uses $\tau D_L$ instead of $\tau$. One can solve for $S(t,\mathbf{r}_l,\{\mathbf{r}_i\})$ for a given initial condition and with known boundary conditions.

When the lamb is \emph{static} ($r=0$), the survival probability $S(t)=[s_1(t)]^N$ \cite{mejia2011first}, where $s_1(t)$ denotes survival probability of the lamb due to a single lion.  Thus, in \emph{closed} geometry $\lim\limits_{t \to \infty}S(t)\sim[\exp(-t/\tau_1)]^N$, and the characteristic time is 
\beq
\label{static_tau}
\tau=\tau_1/N,
\eqn
where $\tau_1$ is the \emph{characteristic} capture time by a single lion. One main focus of this work is to understand how $\tau$ deviates from Eq.~(\ref{static_tau}) as a function of the relative diffusivity ($r$) of the lamb. Before presenting our results, we briefly discuss the computational algorithm for evaluating the survival probabilities.



\section{Numerical Method}
\label{Methods}

\begin{figure}[h!]
\centering
 \includegraphics[width =3.4in]{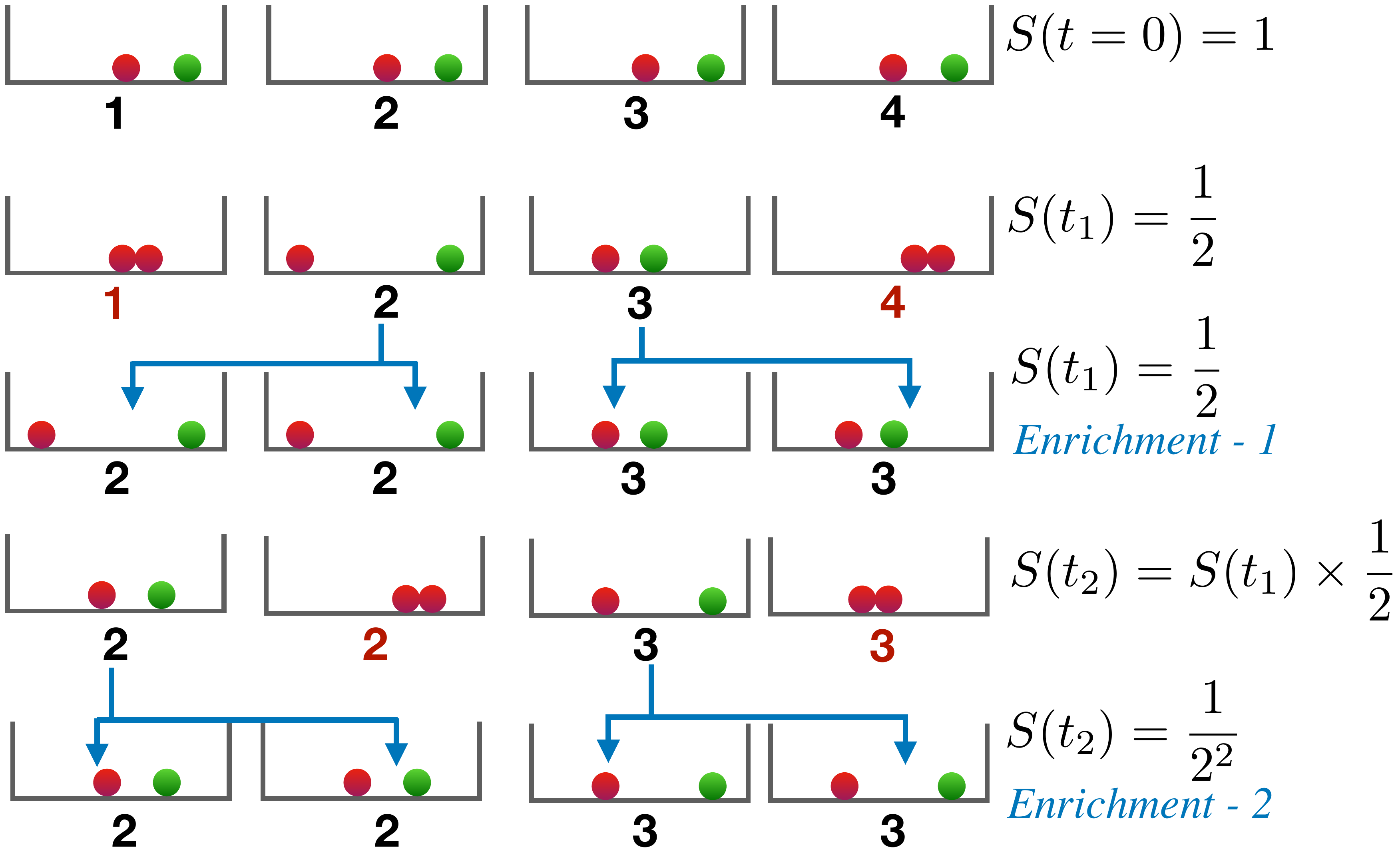}
\caption{(Color online) Illustration of the cloning algorithm to compute the survival probability $S(t)$ of the lamb (red) in the presence of $N$ lions (green) in a $1-d$ setup. At $t=0$, one starts with $M$ realizations of the system. Here $M=4$ and $N=1$. According to this algorithm, when capture happens at time $t_1$ in half of the realizations ($S(t_1)=1/2$), we replace them with the replications of the survived ones (\emph{Enrichment-1}). A similar step (\emph{Enrichment-2}) is performed at time $t_2$ when $S(t_2)=1/4$. We depict the captured lion by changing its color from green to red.}
\label{Fig_grass}
\end{figure}

Obtaining  reliable estimates of $\tau$ from the exponential tail of $S(t)$ is a difficult task as reaching this asymptotic limit numerically is itself challenging. The conventional method of ensemble averaging limits the precision of $S(t)$, which is not good enough to estimate $\tau$. However, based on algorithms proposed earlier in the context of the lamb-lion problem \cite{grassberger2002go}, we have recently shown for a biophysical problem that $S(t)$ can be obtained to very high precision \cite{nayak2020comparison}. The algorithm is depicted for a single lion-lamb pair in Fig.~\ref{Fig_grass}. In general, we start with $M$ realizations of the lion-lamb system at $t=0$. As time progresses, first-passage occurs in some of the realizations, while the remaining number of realizations [say $q(t)$] determines the survival probability $S(t)=q(t)/M$. When $S(t)\leq1/2$, we replace the realizations where capture has occurred with replications of the surviving realizations. This step is called \emph{cloning} or \emph{enrichment} \cite{grassberger2002go}. By doing this, we maintain the ensemble size $M$ constant throughout the simulations. By repeating \emph{enrichment} $n$ times one can obtain the survival probability $S(t)\sim O(1/2^n)$ to a high degree of accuracy. In our simulations, we choose ensemble size $M$ to be  $10^4$ in $1-d$, $3\times10^3$ in $2-d$, and $10^3$  in $3-d$. According to IEEE 754 double-precision floating-point format, the smallest positive number can be stored is $\approx 10^{-308}$ \cite{ieee_85}. Instead of storing very small values of $S(t)$ directly, we always store the logarithm of $S(t)$ at the $n^{th}$ \emph{enrichment} step as $\ln[S(t_n)]=\ln[S(t_{n-1})]+\ln[q/M]$, where $t_n$ represents the time at $n^{th}$ \emph{enrichment} step. The computational cost of generating each $S(t)$ plot varied between few hours to few days.

\begin{figure}[h!]
\centering
\includegraphics[width=3in]{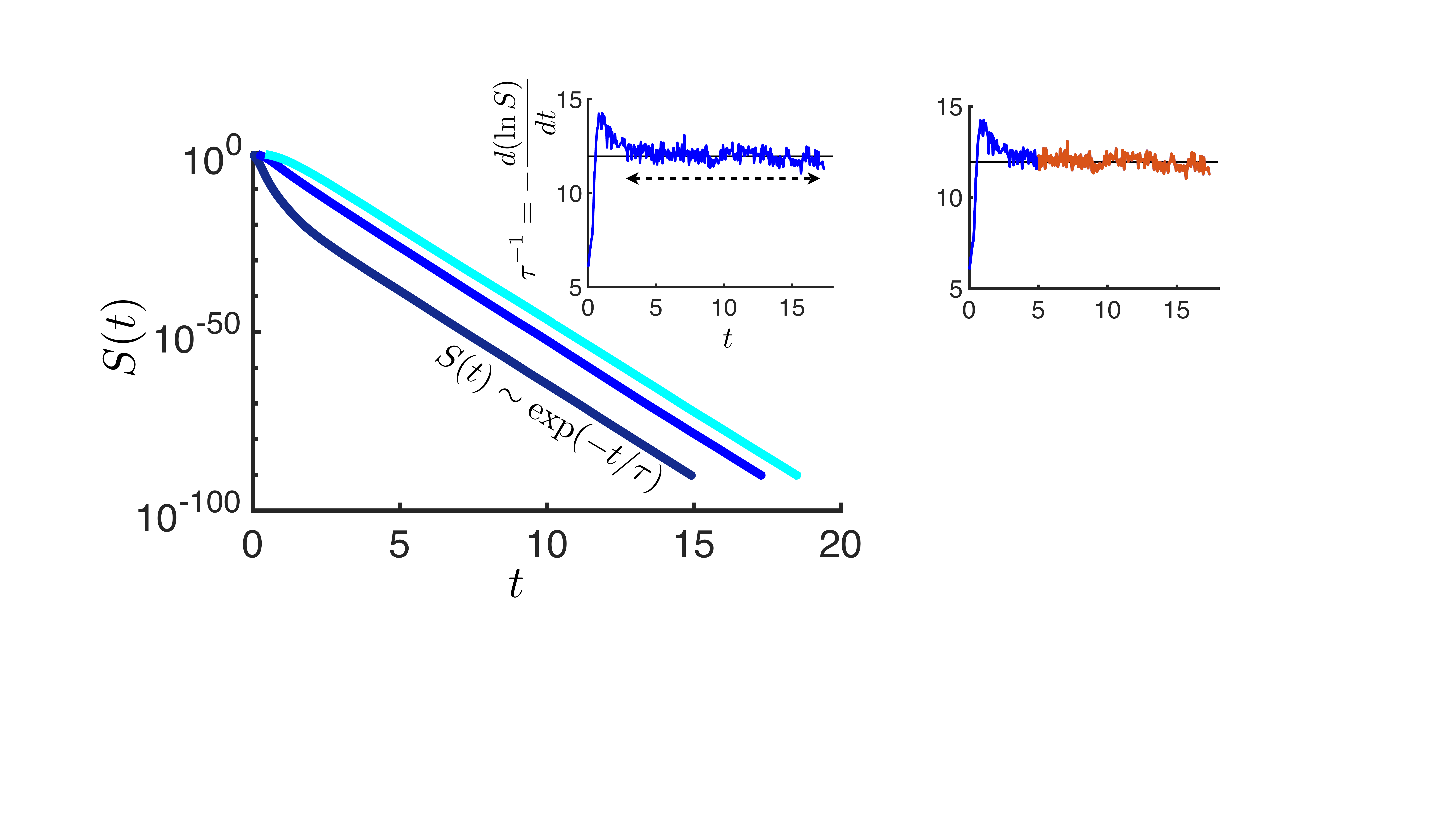} 
\caption{ (Color online) $S(t)$ of a lamb inside a $1-d$ box with $R=4.0$ (arb. unit) and in presence of 50 lions is plotted in a semi-log scale. Three curves correspond to three distinct initial positions $x_0 = 2.5$ (dark blue), 3 (blue), and 3.5 (cyan) (arb. units) of the pack of lions. The initial position of the lamb is $x=1$ (arb. unit) for all the cases.   Although the curves are distinct from each other, asymptotically they are all exponentials with the same $\tau$ -- numerically we obtained $\tau^{-1}=11.98, 11.94$ and 11.79 for the three different cases, respectively, which are very close. A plot of $-d(\ln S)/dt$ vs. $t$ is shown in the inset, and its saturation value $\tau^{-1}$ (in the steady state) is indicated by a black dashed line.}
\label{fig_surv}
\end{figure}


In Fig.~\ref{fig_surv}, $S(t)$ is shown on a semi-logarithmic scale. The three different curves correspond to the three different initial separations between the lamb and the pack of lions. Although the $S(t)$ curves are distinct, note that their asymptotic tails are parallel to each other indicating a unique $\tau$. Moreover for certain initial conditions, we notice that the tail does not even appear for $S(t)\lesssim10^{-30}$. Thus a regular ensemble averaging would have lead to erroneous estimates of $\tau$. By taking negative time derivative of the function $\ln[S(t)]$, we obtain $\tau^{-1}=-\lim\limits_{t \to \infty} \dfrac{d}{dt}[\ln S(t)]$. As shown in the inset of Fig.~\ref{fig_surv}, $\tau^{-1}$ is thus calculated by taking an average of $-\dfrac{d}{dt}[\ln S(t)]$, after it has attained a steady state. Apart from this averaging at the steady-state, the final value of $\tau$ for every case studied in this paper is obtained by further averaging over three different initial conditions. The error bars were within the size of the plotted data points, and therefore not shown here. We observe in our simulation that with increasing $N$, one needs to choose higher values of the enrichment steps $n$ to obtain $\tau^{-1}$ reliably. Depending on $N$, our choice for $n$ varied between $\sim(300-900)$.

\section{Results}
\label{Results}

\subsection{Dependence of $\tau$ as a function of $N$ and $r$ for a moving lamb under confinement}
\label{diff_r}
\begin{figure*}[htbp!]
\centering
\includegraphics[width=7.3in]{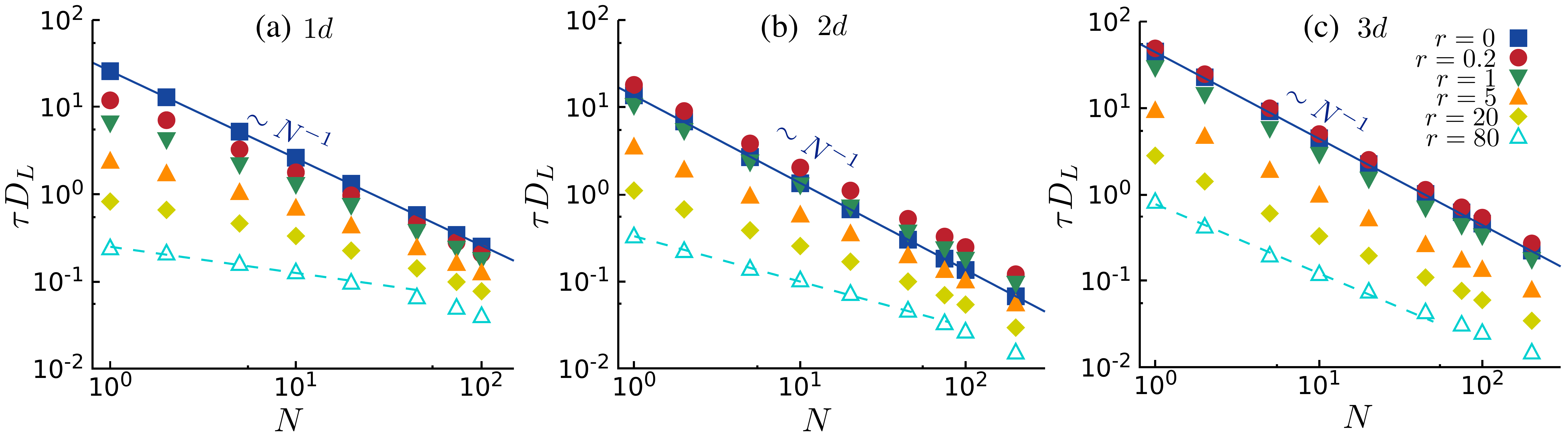} 
\caption{(Color online) $\tau D_L$ vs. $N$ is plotted in the log-log scale for (a) $1d$, (b) $2d$ and (c) $3d$ setups [see Fig.~(\ref{Fig_models})]. Each color of the above figures represents $\tau D_L$ for different ratios $r=0, 0.2, 1, 5, 20, 80$ (see the key) of diffusivities of the lamb and lions ($D_l/D_L$).  \emph{Static} lamb or $r=0$ (navy blue solid square) curves give the exact power-law $\tau D_L \sim N^{-1}$ (navy blue solid line).  For \emph{moving} lamb or $r\neq0$, the curves show  departures from the power-law $ N^{-1}$,  and $\tau D_L$ non-trivially decreases with $N$. At smaller $N$, the power of $N$ changes faster in a lower dimension than the higher dimension upon increasing $r$ (top to bottom). For  $r=80$ (cyan hollow triangle), curves in (a) $1d$ shows $\tau D_L\sim N^{-0.3}$, (b) $2d$ shows $\tau D_L\sim N^{-0.52}$, and (c) $3d$ shows $\tau D_L\sim N^{-0.8}$ at smaller $N$ . Whereas at larger $N$ for different $r$, the curves seem to approach parallel trend by coming closer to each other. The above numerics are done with $a=0.4$ (arb. unit) and $R=8,4,4$ (arb. unit) in $d=1,2,3$ respectively.}
\label{Fig_tau_N}
\end{figure*}
We study how $\tau$ varies with the number $N$ of the lions as well as the relative diffusivity $r$ of the lamb for a fixed $N$, in different dimensions [see Fig.~(\ref{Fig_tau_N})]. Recall when the lamb is \emph{static}, $\tau\sim1/N$  [see Eq.~(\ref{static_tau})] and the exact value of $\tau_1$ is known analytically (see Appendix~\ref{SI_static_L}). We first benchmark our simulations with this limiting results in all the three dimensions. In Fig.~ \ref{Fig_tau_N}(a),(b),(c), we plot $\tau D_L$ for different $N$ and in different dimensions, obtained numerically (blue squares), together with the exact theoretical lines (blue), which shows a very good agreement. The scalings $1/N$ and the prefactor are indeed as expected.


When the lamb is non-stationary ($r\neq0$), there is a clear violation from the simple $1/N$ scaling in all dimensions. Moreover this departure increases upon increasing $r$. A similar deviation from the $1/N$ scaling has been recently reported in context of a biophysical problem, namely the capture of  kinetochore by spindle microtubules \cite{nayak2020comparison}. Here we notice for limited $N$ range ($N\leq100$), $\tau D_L$ roughly follows a power-law form $\sim N^{-\beta}$ with exponent $\beta<1$ and decreasing with increasing $r$. For $r=80$ and at small $N$ range, $\beta \approx 0.3$ in $1-d$ [see cyan dashed line in Fig. ~\ref{Fig_tau_N}(a)]. Note that with increasing $N$, the chances of capture increases, and hence $\tau$ will always be a decreasing function of $N$. However with an increase in the relative diffusivity of the lamb, the chances of a lion-lamb encounter also become higher. This reduces the effectivity of $N$ in regulating the capture time. This could be a possible explanation of why we observe a smaller value of $\beta$ as $r$ increases. A similar trend of $\beta<1$ is also observed in $2-d$ and $3-d$, respectively [Fig.~\ref{Fig_tau_N}(b),(c)]. 


We observe that as the dimensionality increases, the power-law (at small $N$) has lesser departure from the static case. For $r=80$, we notice that the exponent $\beta$ becomes successively bigger --- $0.52$ and 0.8 in $2-d$ and $3-d$ respectively, see the cyan dashed line in Fig.~\ref{Fig_tau_N}(b),(c). Although it is expected that the capture times will be larger for higher $d$ as the effective volume ($R^d/N$) available per lion increases with $d$, it is not immediately clear why at higher dimensions $\tau D_L$ is more sensitive to the variation of $N$. \\


As $N$ becomes large, we saw that the variation of $\tau D_L$ with $N$ starts deviating from the trends discussed above for small $N$. We varied $N$ up to $10^3$, but the $-d/dt[\ln S(t)]$ did not reach steady state similar to the case shown in the inset of Fig.~\ref{fig_surv}. Thus, we were unable to get reliable estimates of $\tau$ beyond $N=200$ even after going down to $S(t)\sim10^{-300}$. We therefore show our data only up to $N=200$. Whether $\tau D_L$ crosses over to a different functional form at large $N$, cannot be ascertained even with the high precision numerics we have. Thus analytical approaches would be preferable to study the $N \to \infty$ limit.

\subsection{The $N$ dependence of $\tau$: a perspective from extreme statistics}
\label{extreme}
In open geometry in $1-d$, the analytical study of the asymptotic dependence of the power-law exponent $\theta_N$ on $N$ associated with $S(t) \sim t^{-\theta_N}$ is based on the statistics of extremes  \cite{redner1999capture}. While the distribution of positions of the non-interacting lions is spreading Gaussians $e^{-x^2/4D_Lt}/\sqrt{4\pi D_L t}$, that of the leader lion closest to the lamb is a Gumbel distribution  \cite{gumbel2012statistics} with a time dependent mean positions $\langle x(t) \rangle=\sqrt{4D_L  t \ln N}$  \cite{krapivsky1996kinetics}. The diffusing lamb thus sees an approaching leader at location $\langle x(t) \rangle \sim \sqrt{t}$ and subsequent analysis leads to  $\theta_N = \ln(N r)/4 r$ \cite{krapivsky1996kinetics,redner1999capture}. 

Under confinement, one may view the problem from two limits. In the first, one may assume that the lion-lamb interaction happens quite fast so that the lions do not see the boundaries of the box by the time capture happens. In that case, like free geometry, the distributions of the lions may be taken as spreading Gaussians, and the leader lion mean position would be  $\langle x(t) \rangle=\sqrt{4D_L  t \ln N}$ 
following Gumbel statistics as discussed above. Assume initial separation of $x_0$ between the lamb and the pack of lions. Given the confinement, the leader only has to travel a finite distance $\sim x_0$ before capture, and hence equating $\langle x(t_c) \rangle \approx x_0$, we get an estimate of a capture time $t_c \approx x_0^2/(4D_L \ln N)$. This prediction of capture times $\sim 1/\ln{N}$ has no similarity with the characteristic times $\tau$ in our numerical study, which go like power-laws in $N$. 

A second view is that characteristic times represent rare and long-lived events, and by the time a capture happens, the lions explore  
a substantial part of the available finite volume ($\sim R^d$) and have many reflections off the boundary walls. Thus in this limit, we may assume that the probability distribution of the radial position $r_i$ (assuming radial symmetry) of the $i^{th}$ lion may be assumed to be  uniform, i.e. $P(r_i,t) \approx C_d$ over a sub-volume of order $\sim R^d$.  Statistics of extremes predict the cumulative distribution of the position $w$ of the leader lion closest to the lamb to be given by (for large $N$): 
\beqr
Q_L(w) &=& Q_L(w <{\min}\{r_i\}) \nonumber \\
&=& {\rm Prob}(w < r_1) {\rm Prob} (w < r_2) \ldots {\rm Prob}(w < r_N) \nonumber \\
&=& \big[1 - \int_0^{w} P(r_i) \Omega_d r_i^{d-1} dr_i\big]^N \nonumber \\
&=& \exp(-q_d N w^d), \nonumber
\eqnr
and the corresponding probability distribution is the Weibull distribution  \cite{kotz2000extreme,rinne2008weibull}
\beq
P_L(w) = - [dQ_L(w)/dw ]=  q_d N w^{d-1} d \exp(-q_d N w^d). 
\eqn
Here $\Omega_d$ is the $d$-dimensional solid angle and $q_d=\Omega_d C_d/d $. The peak value of $P_L(w)$ occurs at 
$w_*=\Big[\frac{(d-1)}{d q_d N}\Big]^{1/d}$ (for $d>1$). While for $d=1$ a typical position would be $w_1 \sim 1/{N}$. Now if we 
assume that the lamb travels diffusively within this length scale $w_*$ (or $w_1$ for $1d$) and gets captured, then an estimate of capture time: 
\begin{align}
t_c \sim w_*^2/D_l \sim N^{-2/d}
\end{align}
in general $d$.  It is interesting that this Weibull distribution based view gives at least a power-law capture time as we observe for our numerical estimates of $\tau$ for finite $N$. But the values of the predicted exponent $\beta=2/d$ ($=2, 1$ and $2/3$ in $1,2$, and $3$-dimensions respectively) are unrelated to what we find in our numerical study.  Moreover, $\beta$ has no $r$ dependence, as we find in our simulations.  Thus this approximate analytical argument is not appropriate to explain the accurate computational data but is indicative that power-laws may arise at large values of $N$.
 
\subsection{Comparison with previous analytical study for $N=1$}
\label{compare_apprx}
\begin{figure}[!htbp]
\centering
\includegraphics[width=3in]{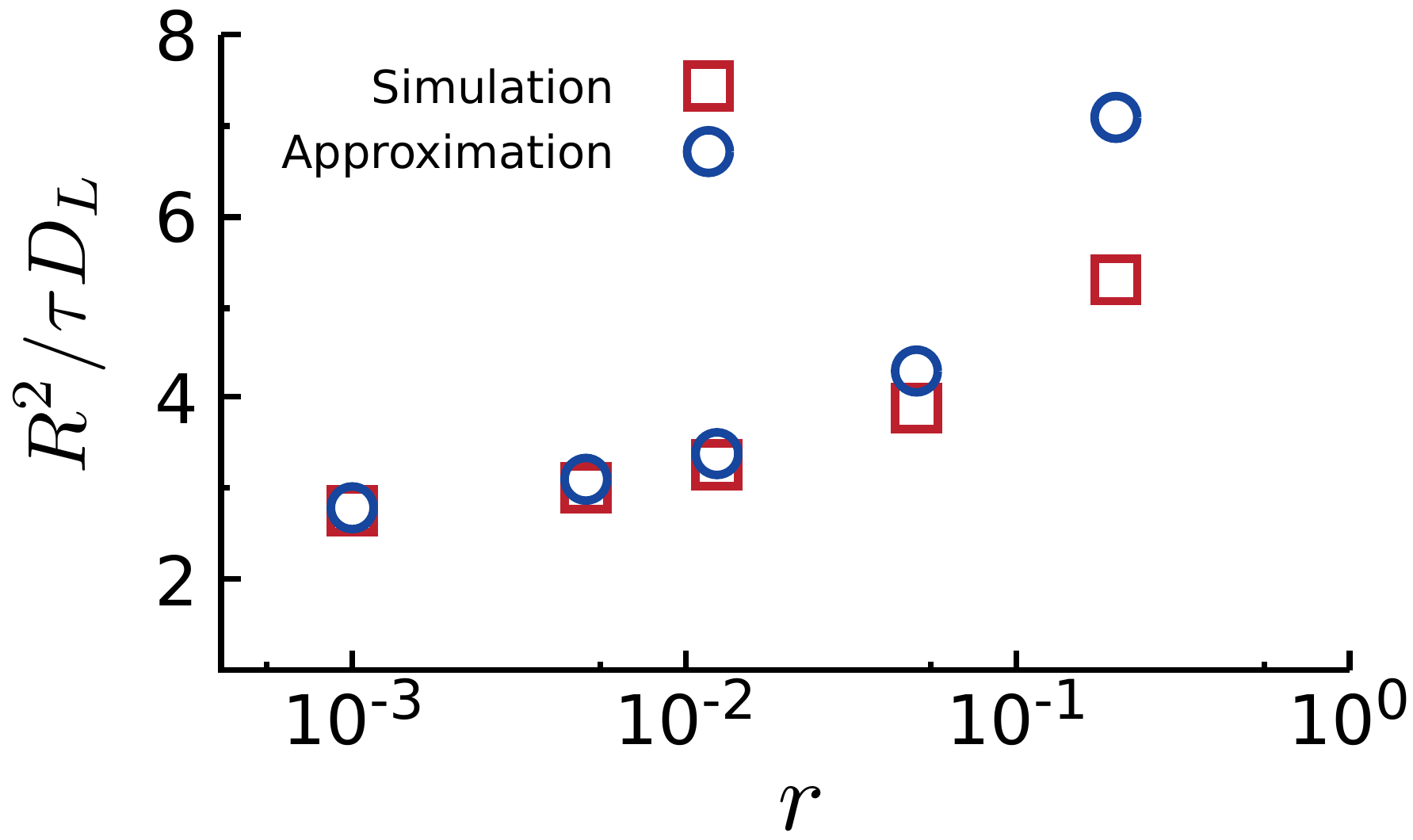} 
\caption{Comparison of $R^2/\tau D_L$ obtained numerically (red squares) with the approximate analytical prediction in Eq.~(\ref{Eq_vot}) (blue circles) for different values of $r$.}
\label{Fig_vot}
\end{figure}
As mentioned earlier, the analytical solution of $S(t)$ is known for a couple of cases for a single lamb-lion pair ($N=1$) in 1\emph{d}. When $D_l=D_L=D$, it is known exactly that $\tau=R^2/\pi^2 D$ \cite{tejedor2011encounter}. For $r\neq1$, the problem is not solvable exactly. An approximate answer is known in the limit of $r \to 0$~$(D_l \ll D_L)$ \cite{vot2020first}:
\beq
\begin{aligned}
\tau (D_l,D_L) \simeq \frac{4R^2}{\pi^2D_L D_l} (D_l+D_L)\arctan ^2(\sqrt{D_l/D_L})\\
\times \Big(1+ \frac{2^{4/3}\alpha'}{\pi^{2/3}} \arctan ^{2/3}(\sqrt{D_l/D_L} \Big),
\end{aligned}
\label{Eq_vot}
\eqn
where $\alpha'\approx-1.0188$ is the first zero of the derivative of Airy function. We compare our high precision numerical estimates of the timescales with this analytical result (see Fig.~\ref{Fig_vot}). We see that the analytical approximation of Ref.~\cite{vot2020first} agrees with our numerical results for $D_l/D_L \lesssim 0.01$.

\subsection{Scaled characteristic time $\tau D_L/g_d(R,a)$ depends mainly on $r$ and $N$} 
\label{diff_R}

\begin{figure*}[htbp!]
\centering
\includegraphics[width=7.2 in]{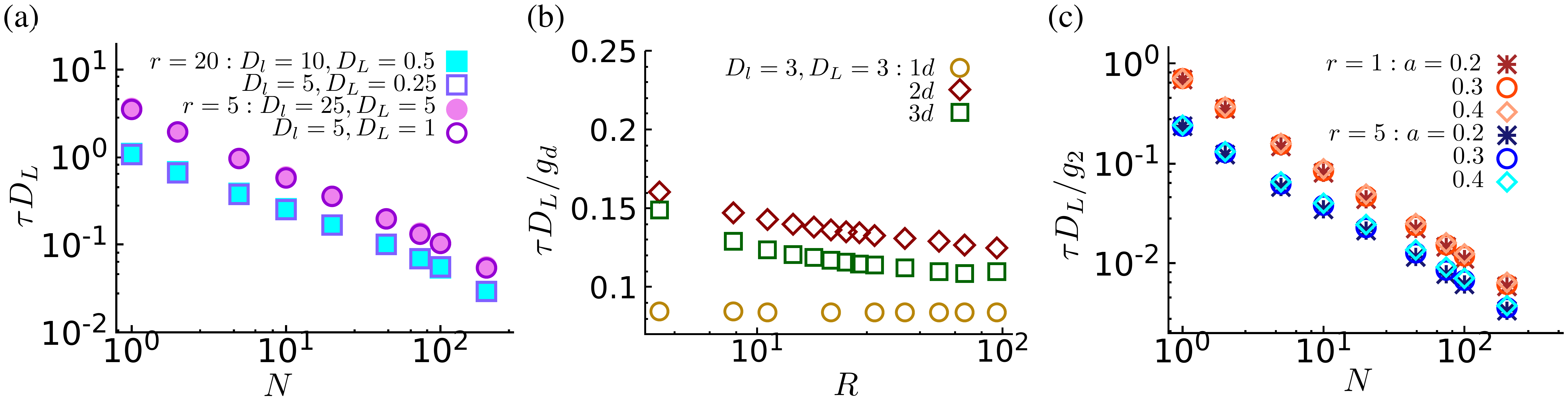} 
\caption{ (Color online) (a) $\tau D_L$ vs. $N$ is plotted for $2d$ setup [see Fig.~\ref{Fig_models}(b)] for $r=5, 20$ in log-log scale. $\tau D_L$ is a function of $r$ and does not depend upon the individual diffusivities $D_l$ or $D_L$.  For a fixed $r$, $\tau D_L$ collapses for any pair of diffusivities $D_l, D_L$.
 (b) $\tau D_L / g_d$ vs. $R$ is plotted for $r=1$ with fixed $a=0.4, N=5$ in $d=1,2,3$.  Where, $g_d(R,a)$  corresponds to the scaling of $\tau D_L$  for the \emph{static} lamb case, in $d$-dimension. (c) $\tau D_L/g_2$ vs. $N$ is plotted for three distinct radii  $a= 0.2,~0.3, ~0.4$ of the lamb,  in log-log scale  with fixed $R=0.4$. We show this for two different relative diffusivities $r=1 (D_l=3,D_L=3), 5(D_l=5, D_L=1)$ [see Fig.~\ref{Fig_models}(b) setup]. }
\label{Fig_diff_L}
\end{figure*}

For a first passage process, the timescales of capture will in general depend on the system parameters. For example in our study, $\tau=f(D_l, D_L,N,R,a)$ in general. However, $\tau$ can be scaled appropriately such that the scaled quantity depends mainly on $N$ and $r$, as we show below. This is also the main reason for presenting the $N$ and $r$ dependent study of $\tau$ separately in Sec.~\ref{diff_r}.  

As shown earlier, if we scale time with $D_L$ [see Eq.~(\ref{BFP})], then the results will only depend on the relative diffusivity $r$ and the other parameters, \ie ~$\tau D_L=f_1(r,N,R,a)$. A verification of this important feature is shown in Fig.~\ref{Fig_diff_L}(a) for $d=2$ with fixed $R$ and $a$. For different sets of lamb-lion diffusivities $(D_l,D_L)$ but with equal relative diffusivity $r$, the curves of $\tau D_L$ versus $N$ collapse. 


Next, we look at the dependence of $\tau D_L$ on the system dimensions, $R$ and $a$. Since we do not have analytical expressions for $f_1$, we  proceed numerically. Additionally, we start with an initial guess (although not guided by any theory) that these dependences are same as in the case of a static lamb. Our numerical results show below that this is a  good approximation for (reasonably) large system sizes $R$. In Appendix \ref{SI_static_L}, the derivations of the dependence of $\tau$  on the $R$ and $a$ for the case of a \emph{static} lamb in $d=1,2,3$ are shown. These results are as follows:
\beqr
&1-d:&~~~\tau D_L=\Big{(}\frac{4 R^2}{\pi^2}\Big{)}\frac{1}{N}, \nonumber \\
&2-d:&~~~\tau D_L \approx \Big{(}\frac{R^2}{2} \ln[(R/a)-0.5]\Big{)}\frac{1}{N}, \nonumber \\
&3-d:&~~~\tau D_L \approx \Big{(}\frac{[R-a]^3}{3a}\Big{)}\frac{1}{N}.
\label{len}
\eqnr
In Eq.~(\ref{len}), the terms in the parenthesis, which we will denote as $g_d(R,a)$, indicate the dependence on $R$ and $a$. Note that for $1-d$, the lamb is taken to be a point-sized particle ($a=0$). In $2-d$ and $3-d$, the relationships are approximate and hold true for $R\gg a$ (see the exact transcendental equations in Appendix A). For a static lamb, the $1/N$ scaling is true for every dimension. As shown earlier in Sec.~\ref{diff_r}, for a moving lamb, there is a deviation from this $1/N$ behavior. Yet we find that the prefactor $g_d$ remains roughly unchanged if we compare the moving and the static cases.  To show this, we first plot $\tau D_L/g_d$ versus $R$ in Fig.~\ref{Fig_diff_L}(b) for fixed $r \neq 0$, $N$, and $a$. We see that for $d = 2,3$, the curves are essentially constant at large $R$, while for $d=1$, the curve is constant at any $R$. This suggests that the approximation is good even in the case of a moving lamb. Thus the function $f_1(r,N,R,a) \approx g_d(R,a)f_2(r,N)$. Next, to show that there is no strong dependence of $f_2$ on $a$, we plot $\tau D_L/g_2$ versus $N$ in $2d$ for two different values of $r$ with three distinct values of $a$ in each case. In Fig.~\ref{Fig_diff_L}(c), for both values of $r$, the data collapse for different radii $a$, confirming the hypothesis.   

\section{Conclusion}
\label{conclusions}
The capture of a moving target by multiple random walkers is an important problem in stochastic processes and has been extensively studied before in free-space. However a similar study in confined space is still lacking. The presence of confinement is pertinent to many physical and biological processes. Such a constraint makes the study challenging due to nonlinear effects introduced by the presence of boundaries. In fact, so far no exact result exists for the problem of moving lamb chased by multiple lions in confined geometry, except for a single lion-lamb pair, that too in $1$-dimension. In this work, we took a computational approach to study this classic open problem for multiple lions and in $1,2$, and $3$ dimensions. We did an extensive characterization of the capture process by accurately computing the \emph{characteristic} capture timescale $\tau$ (which is an initial position-independent quantity), and studied its dependence on the various system parameters. The significant point is that the estimates are based on very high precision computation of $S(t)$ (to $\sim 10^{-300}$). Hence the results would serve as reliable checkpoints for future  analytical theories, either approximate or exact. \\
 
Here the main system parameters are the number of lions $N$ and  the relative diffusivity $r$. We showed that instead of the characteristic time itself, a more relevant quantity to study is the scaled quantity $\tau D_L$, which depends only on the relative diffusivities $r$. We first benchmarked our numerical estimate of the characteristic time by computing it for the static case ($r=0$) with varying $N$ in all the dimensions. We found a very good agreement of our result to the $\sim 1/N$ scaling, which is known theoretically. For the case of non-stationary lamb ($r \neq 0$), 
our results exhibit a clear deviation from the $1/N$ scaling. We noticed that the timescale varies non-trivially with $N$, which cannot be explained by a single power-law scaling and depends on the relative diffusivity $r$ as well the dimension-$d$ of the system. \\

Our computational study of obtaining $\tau$ even though of high precision, is inadequate for large $N > 200$. Obtaining the asymptotic exponential tail of $S(t)$ becomes increasingly difficult, and computationally expensive leading to unreliable estimates of $\tau$. Using heuristic arguments based on the statistics of extremes, we speculated an $N$ dependence of $\tau$.  The theoretical prediction is $\tau \sim N^{-2/d}$ at large $N$, indicating the possibility of a power-law behavior. Although not quantitatively consistent with our numerical results, which was obtained over a limited range of $N$, this new possibility based on Weibull statistics may be of interest to problems within confined geometries.  
\\

Interestingly we notice that the dependence of the scaled quantity $\tau D_L$ on 
system dimensions $R$ and $a$ is quite similar (for large $R$) to that of the static case, 
which is known theoretically. We numerically verified this in all dimensions by appropriately scaling $\tau D_L$ by the relevant function $g_d(R,a)$. 
 \\
 
Our central result showing the deviation of the characteristic time from the $1/N$ scaling for non zero diffusivity of the lamb may be of general interest in problems of biophysics and ecology. Recently similar power-laws with exponent that differs from unity, was reported for the problem of a kinetochore capture by mobile microtubules within a nuclear volume \cite{nayak2020comparison}.  For approximate analytical studies, our results may be used to test the goodness of the approximation. We made such a comparison to demonstrate the validity of a known approximate result for a lion-lamb pair in $1$-dimension with unequal diffusivities. We hope our study will revive interest in this rather general first passage problem, which may arise in various phenomena in physical and biological sciences. \\


\section{Acknowledgements}
 The authours thank Satya N. Majumdar for fruitful discussions. D.D. acknowledges SERB India (grant no. MTR/2019/000341) for financial support. A.N. acknowledges IRCC at IIT Bombay, India, and SERB, India (Project No. ECR/2016/001967), for financial support. I.N. thanks IIT Bombay for  the Institute Ph.D. fellowship, and acknowledges the High-Performance Computing Facility at IIT Bombay. 

\appendix

\section{Dependence of $\tau$ on the system size $R$ and $a$ for the \emph{static} lamb}
\label{SI_static_L}
One may find the dependence of $\tau$  on the system size $R$ for a \emph{static} lamb, from the backward Fokker-Planck equation for survival probability  [see Eq.~(\ref{BFP})]. Since for the \emph{static} lamb in a bounded domain, the \emph{characteristic} capture times in case of  $N$  lions are always $\tau=\tau_1/N$, it is  sufficient to calculate the dependence of $\tau_1$ (the timescale for $N=1$ lion)  on $R$ and $a$. Substituting $D_l=0$, $N=1$ in Eq.~(\ref{BFP}), we get 
\begin{equation*}
\frac{\partial S(t,\mathbf{r}_1)}{\partial t}= D_L\nabla ^2 S(t,\mathbf{r}_1).
\end{equation*}
Where $\nabla_1 ^2 =\frac{d^2}{dr_1^2}+\frac{(d-1)}{r} \frac{d}{dr_1}$ is the $d-$dimensional radially symmetric Laplacian operator. For all the spatial dimensions, we consider the lamb to be at the origin [see Fig.~\ref{Fig_models}]. By applying spherically symmetric \emph{absorbing boundary} condition : $S(r_1= a)=0$ ($a$ is 0 for $d=1$), and \emph{reflecting boundary} condition : $\frac{\partial S}{\partial r_1}\mid_{r_1 = R}=0$ for $d=1,2,3$, we may obtain solutions of the form: $S(r_1,t)=\sum_{k}c_{k} \tilde R_{k}(r_1)e^{-k^2D_Lt}$. Note that $c_k$ is a constant that obeys $\sum_{k}c_{k} \tilde R_{k}=1$ due to the {\it initial condition} $S(r_1,0) = 1$. The time-independent function $ \tilde R_k(r_1)$  represents a $d$-dimensional radial solution, with the modes $k$ satisfying the following transcendental equations: 
\begin{equation}
\begin{aligned}
1d: k=(2n+1)\pi/2R,  \\
2d: J_1(kR)Y_0(ka)=J_0(ka)Y_1(kR),\\
3d : \tan[k(R-a)]=kR.\\
\end{aligned}
\label{trans}
\end{equation}
Here for $1d$, integer $n=0,1,...,\infty$. For $2d$, $J_i$ and $Y_i$ represent the Bessel functions of $i^{th}$ order for first and second kind. We note that $\tau_1$  corresponds to the longest timescale which further corresponds to the smallest value of $\{k\}$ (say $k_{\min}=\min\{k\}$)  in  Eq.~(\ref{trans}) such that: $\tau_1=1/(k_{\min}^2 D_L)$.

\subsection{Length dependence for $1d$} 
By substituting $n=0$ in Eq.~(\ref{trans}), we get the exact expression of $\tau_1$ as a function of $R$ for $1d$ :
\begin{equation}
\begin{aligned}
\ k_{\min}^2=[\tau_1 D_L]^{-1}=\frac{\pi^2}{4R^2}\\
\Rightarrow 
\tau_1 =\frac{4 R^2}{\pi^2 D_L}.
\end{aligned}
\label{SI_1d}
\end{equation}

\subsection{Length dependence for $2d$} 
We approximate value of $k_{\min}$ in Eq.~(\ref{trans}) by
expanding $J_0,J_1, Y_0,Y_1$ at small $k$ limit.  Expansions of $J_0 (x),J_1(x), Y_0(x),Y_1$ are as follows: 
\begin{align*}
J_0(x)&=1-\frac{x^2}{2}+\frac{x^4}{64}+O[x^6]\\
 J_1(x)&=\frac{x}{2}-\frac{x^3}{16}+\frac{x^5}{384}+O[x^6]\\
 Y_0(x)&= \frac{2}{\pi}\bigg[\ln(\frac{x}{2})+\gamma\bigg]+\frac{x^2}{2\pi}\bigg[1-\gamma+\ln(\frac{2}{x})\bigg]+O[x^3]\\
 Y_1(x)&=-\frac{2}{\pi x}+\frac{x}{2\pi}\bigg[2\gamma-1+\ln(\frac{x^2}{4})\bigg]+O[x^3].
\end{align*}

Here $\gamma$ is the Euler-Mascheroni constant. Retaining up to linear order 
analytic terms in $x=k_{\min}$, plus the singular terms, we get, 
 \begin{eqnarray*}
J_0(k_{\min} a) &\approx& 1, \\
 J_1(k_{\min} R) &\approx& \frac{k_{\min} R}{2}, \\
 Y_0(k_{\min} a) &\approx& \frac{2}{\pi}\bigg[\ln\bigg(\frac{k_{\min} a}{2}\bigg)+\gamma\bigg], \\
 Y_1(k_{\min} R) &\approx& -\frac{2}{\pi k_{\min} R} \\
 && +\frac{k_{\min} R}{2\pi}\bigg[2\gamma-1+\ln\bigg(\frac{(k_{\min} R)^2}{4}\bigg)\bigg]. 
 \end{eqnarray*}
 
 By substituting the above equations in Eq.~(\ref{trans}) we get,
\begin{equation}
\begin{aligned}
k_{\min}^2\approx\frac{2}{R^2(\ln(\frac{R}{a})-0.5)}\\
\Rightarrow \tau_1 D_L\approx \frac{R^2}{2}(\ln(\frac{R}{a})-0.5).
\end{aligned}
\end{equation} 
 \vspace{1mm}
 
\subsection{Length dependence for $3d$}
The Taylor series expansion of $\tan(x)$ about $x=0$ is

\begin{equation*}
\tan(x)=x+\frac{x^3}{3}+ O[x^5]
\end{equation*}
Expanding Eq.~(\ref{trans}) for small $k$ and keeping up to cubic term we get,
\begin{equation}
\begin{aligned}
k_{\min}(R-a)+\frac{k_{\min}^3(R-a)^3}{3} \approx k_{\min} R\\
k_{\min}^2\approx \frac{3a}{(R-a)^3}\\
\Rightarrow \tau_1 D_L\approx \frac{(R-a)^3}{3a}.
\end{aligned}
\eqn

\bibliography{v1}

\end{document}